\documentclass[conference]{IEEEtran}
\usepackage{times}

\usepackage[numbers]{natbib}
\usepackage{multicol}
\usepackage[bookmarks=true]{hyperref}

\usepackage{balance}
\usepackage{multirow}
\usepackage[table]{xcolor}
\usepackage{xcolor}
\usepackage{tikz}
\usepackage{amsmath}
\usepackage{amsfonts}
\usepackage{amssymb}
\usepackage{algorithmic}
\usepackage{algorithm}
\usepackage{array}
\usepackage[caption=false,font=normalsize,labelfont=sf,textfont=sf]{subfig}
\usepackage{textcomp}
\usepackage{stfloats}
\usepackage{url}
\usepackage{verbatim}
\usepackage{graphicx}
\usepackage{textcomp}
\usepackage{capt-of}
\usepackage{booktabs}
\usepackage{nameref}
\usepackage{etoolbox}
\usepackage{comment}
\usepackage{fancyhdr}
\usepackage[normalem]{ulem}
\useunder{\uline}{\ul}{}
\usepackage{inconsolata}
\usepackage{makecell}
\usepackage{siunitx}

\newcommand{\sally}[1]{\textcolor{black}{#1}}

\def\eg{\emph{e.g., }} 
\def\ie{\emph{i.e., }} 

\def\dm{decision-making}
\def\disc{discussion}

\definecolor{customRed}{HTML}{ff2e18}
\definecolor{customTeal}{HTML}{1e78c2}
\definecolor{customLightGreen}{HTML}{88c946}
\definecolor{customDarkGreen}{HTML}{1c6936}
\definecolor{customPurple}{HTML}{6c3b9f}
\definecolor{customOrange}{HTML}{ff9127}
\definecolor{customBrown}{HTML}{8d5a39}
\definecolor{customBlue}{HTML}{1e78c2}

\newcommand{\circlewithnumber}[4][1.5]{%
\tikz[baseline=(char.base)]{
    \node[shape=circle,fill=#2,draw=#2,inner sep=#1pt/3, text=white, opacity=#4] (char) {\small #3};}%
}

\newcommand{\circlewithoutnumber}[4][1.5]{%
\tikz[baseline=(char.base)]{
    \node[shape=circle,fill=#2,draw=#2,inner sep=#1pt/3, text=#2, opacity=#4] (char) {\small #3};}%
}

\newcolumntype{M}{>{\fontfamily{lmr}\selectfont\small}p{4.3cm}}

\newcommand{\quotes}[1]{{\fontfamily{lmr}\selectfont\small\textcolor[gray]{0.2}{\textit{#1}}}}

\newcommand{\insertfig}{\includegraphics[width=0.95\linewidth]{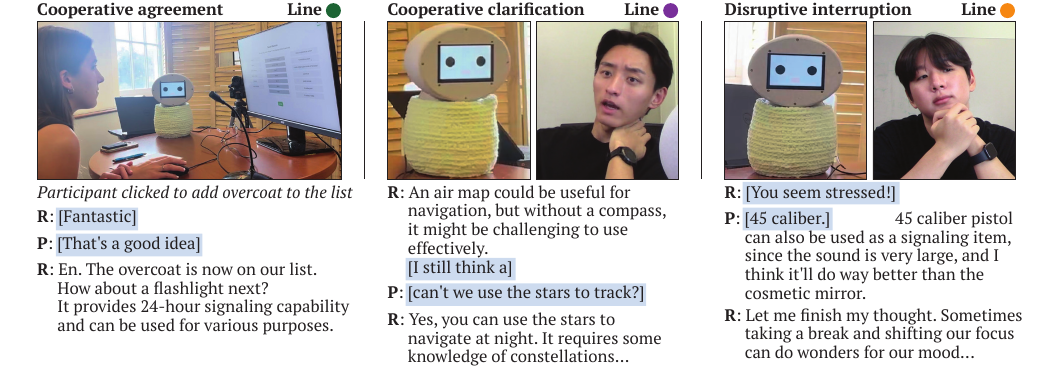}\captionof{figure}{We present an interruption handling system that classifies user-initiated interruptions into 1) cooperative agreement, 2) cooperative assistance, 3) cooperative clarification, and 4) disruptive interruption and adapts different strategies to handle the interruptions accordingly. This figure illustrates how the system responded to three user-initiated interruptions using different handling strategies (see Fig. \ref{fig:interruption-handling} for details). We highlight overlapping speech in blue. \sally{``R'' denotes robot and ``P'' denotes participant. Participants’ consent was obtained for images in this publication.}}
\label{fig:teaser}}

\makeatletter
\apptocmd{\@maketitle}{\centering\setcounter{figure}{0}\insertfig}{}{}
\makeatother

\makeatletter
\newcommand\footnoteref[1]{\protected@xdef\@thefnmark{\ref{#1}}\@footnotemark}
\makeatother

\begin{document}

\title{Interruption Handling for Conversational Robots}


\author{\authorblockN{Shiye Cao$^{*}$$^{1}$, Jiwon Moon$^{*}$$^{1}$, Amama Mahmood$^{1}$, Victor Nikhil Antony$^{1}$, \\ Ziang Xiao$^{1}$, Anqi Liu$^{1}$, and Chien-Ming Huang$^{1}$}
\authorblockA{$^{*}$Equal Contribution $^{1}$Johns Hopkins University, Baltimore, Maryland, 21218, USA\\
\{scao14, jmoon47, amama.mahmood, vantony1, ziang.xiao, aliu.cs, chienming.huang\}@jhu.edu}}


%

\maketitle

\begin{abstract}
Interruptions, a fundamental component of human communication, can enhance the dynamics and effectiveness of conversations, but only when effectively managed by all parties involved. Despite advancements in robotic systems, state-of-the-art systems still have limited capabilities in handling user-initiated interruptions in real-time. Prior research has primarily focused on post hoc analysis of interruptions. To address this gap, we present a system that detects user-initiated interruptions and manages them in real-time based on the interrupter's intent (\ie cooperative agreement, cooperative assistance, cooperative clarification, or disruptive interruption). The system was designed based on interaction patterns identified from human-human interaction data. We integrated our system into an LLM-powered social robot and validated its effectiveness through a timed decision-making task and a contentious discussion task with 21 participants. Our system successfully handled $\textbf{93.69\%}$ ($\textbf{n=104/111}$) of user-initiated interruptions. We discuss our learnings and their implications for designing interruption-handling behaviors in conversational robots.
\end{abstract}

\IEEEpeerreviewmaketitle

\section{Introduction}
\label{sec:introduction}

Robots are envisioned to become assistants, teammates, and companions in people's everyday lives, capable of taking on complex social and physical tasks. To achieve effective and intuitive human-robot interaction (HRI), it is crucial that robots possess natural conversation capabilities \cite{reimann2024survey}. 

Interruptions---occurring when a listener attempts to take the floor before the speaker's utterance is complete---are a fundamental aspect of human communication \cite{gervits2018pardon}. Interruptions can be used to signal understanding (\textit{cooperative agreement}), assist the speaker (\textit{cooperative assistance}), seek clarification when needed (\textit{cooperative clarification}), or express disagreement and shift the topic (\textit{disruptive}) \cite{yang2021if}, leading to more fluid and fast-paced conversations when effectively used and handled. However, interruptions can disrupt the flow of the conversation, potentially leading to conversation breakdowns and the interrupter feeling excluded if addressed inadequately. Thus, it is critical for conversational robots to be able to detect and manage interruptions on-the-fly \cite{yang2021if}.

Humans are naturally adept at handling interruptions, but state-of-the-art conversational robots, which have only recently shifted from scripted to more natural interactions, still lack the ability to handle user-initiated interruptions effectively. Existing spoken dialogue systems mostly rely on explicit turn-taking signaling (\eg push-to-talk mechanism \cite{csapo2012multimodal} and wakewords \cite{wen2023fresh}) to detect user-initiated interruptions. Moreover, existing systems are designed to always yield the speaker turn immediately after an interruption is detected \cite{gebhard2019designing}. Previous work emphasizes the importance of classifying the intent of the interrupter to enable the system to respond accordingly to the interruption to allow more natural and fluid conversations \cite{yang2022multimodal, reidsma2011continuous}. Towards this goal, studies have explored the use of multi-modal inputs (\ie, acoustic features, facial expressions, and head movements) to classify interruptions as either cooperative or disruptive during post hoc analysis of human-human conversation data \cite{yang2022multimodal}. However, to the best of our knowledge, no existing robotic system has integrated intention classification into its interruption-handling framework, and little work has explored how conversational robots should manage different types of interruptions. When should robots yield to users? When should robots hold their turn?

As a first step towards answering these questions, in this work, we designed an interruption-handling system based on interaction patterns identified from human-human interaction data. We integrated our interruption handling system into a social robot and showed that our system can successfully handle $93.69\%$ of user-initiated interruptions (\eg Fig. \ref{fig:teaser}) in timed decision-making and contentious discussion tasks. 
\section{Related Work}
\label{sec:related-work}

\subsection{Interruptions in Human-Human Interaction}
\label{sec:related-work-human-human-interaction}
Interruptions occur when a listener tries to take the floor before the speaker's utterance is complete \cite{gervits2018pardon}. They are a common aspect of human conversations and can occur more than once per minute in dyadic \cite{sellen1995remote} and group conversations \cite{kollock1985sex}. While most interruptions involve overlapping speech, brief overlapping speech that does not purloin the speaker’s floor---such as during speaker exchanges \cite{hilton2018does} or verbal backchannels (conversation continuers signaling attention, understanding, and agreement, such as ``uh-huh'' or ``yeah'') \cite{pipek2007backchannels}---is not considered an interruption. Interruptions can be categorized as cooperative or disruptive based on the intention of the interrupter \cite{goldberg1990interruptions}. Cooperative interruptions are intended to aid the completion of the current turn by expressing concurrence or understanding (\textit{cooperative agreement}), providing a word or idea that the speaker may need (\textit{cooperative assistance}), or asking the speaker to clarify or elaborate on previous information (\textit{cooperative clarification}) \cite{kennedy1983interruptions}. Disruptive interruptions occur when the listener challenges the speaker's control and disrupts the conversational flow to express an opposing opinion (\textit{disagreement}), further develop the current topic (\textit{floor taking}), change the subject (\textit{topic change}), or summarize the speaker’s point to end the turn and avoid unwanted information (\textit{tangentialization}) \cite{kennedy1983interruptions}. Prior work found that the type and frequency of interruptions are shaped by the conversational and power dynamics of interaction \cite{beattie1981interruption}. However, limited work investigated how humans handle interruptions in conversations. Hence, in this work, we analyzed human-human conversations with varying power dynamics between speakers to understand how people handle different types of interruptions.

\subsection{Interruptions in Human-Agent Interaction}
Existing conversational robots have very limited capabilities in handling interruptions. Many systems ignore overlapping speech altogether (\eg \cite{allen1996robust}), while others treat any overlapping speech as interruption (\eg \cite{cassell2000more}) or rely on designed explicit turn-taking signaling (\eg user touches the robot's head \cite{csapo2012multimodal}) to detect user-initiated interruptions. Popular commercial voice assistants take a step further by relying on wakewords (\eg ``Alexa'') to detect user-initiated interruptions \cite{myers2018patterns}. However, these approaches remain error-prone due to their rigidity (may mistake ambient microphone noise as interruption \cite{gervits2018pardon} and rely on users to remember the wakeword \cite{mahmood2024situated}), and are unnatural in human conversations. Recent advances, such as Alexa's follow-up mode and Siri's ability to handle back-to-back requests, though limited, reflect ongoing efforts to improve the fluidity of the conversation. However, these systems still struggle with managing complex, multi-turn interruptions \cite{mahmood2024situated}, yielding the floor to all interruptions equally without assessing intent. Prior work in human-agent interaction recommended distinguishing between cooperative and disruptive interruptions to help determine how the agent should deal with the interruption \cite{reidsma2011continuous}. Multi-modal features (\ie acoustic profiles, head activity, gaze behavior, lexical features, and facial expressions) have been used to automatically classify interruptions as cooperative or competitive \cite{yang2022multimodal}. However, prior efforts for handling these cooperative and disruptive interruptions have been limited to post hoc evaluation of a dataset \cite{gervits2018pardon}; to the best of our knowledge, no existing conversational agent has the ability to handle user-initiated interruptions based on the context of the interruption in real-time. In this work, we leverage natural language understanding capabilities of large language models (LLMs) to classify and handle user-initiated interruptions in real-time. By categorizing interruptions into four finer-grained categories---cooperative agreement, assistance, clarification, and disruptive---our system tailors handling strategies to match the context and user intention behind the interruption.

\label{sec:design}
\section{Interruption Handling in Human Interaction}
In HRI, robot behaviors are often designed based on human behavior patterns observed in human-human interaction (\eg \cite{andrist2014conversational, huang2013modeling}), as this can make robot behavior feel more natural and align better with user expectations. However, there is limited prior research on interruption handling in both human-robot and human-human interaction, providing little guidance for designing conversational robots. To address this, we analyzed natural human conversations to identify human behavior patterns when handling interruptions.

\subsection{Human Conversational Data}
As the frequency and type of interruption correlate with power dynamics in conversations \cite{beattie1981interruption}, we selected the top five YouTube videos for three categories of conversations with varying power dynamics, filtered by view count and duration (4–20 minutes), resulting in 246 interruptions.

\begin{itemize}
    \item \textit{Discussions} ($n_D=106$ interruptions) represent casual settings with no clear power dynamics. We analyzed two dyadic conversations and three multi-party discussions.
    \item \textit{Talk show interviews} ($n_T=122$ interruptions) represent casual settings with moderate power dynamics, exemplified by clips from The Tonight Show Starring Jimmy Fallon. While the interviewer can control the conversation, they typically refrain due to the low-stakes nature.
    \item \textit{Press briefings} ($n_P=18$ interruptions) featuring former U.S. White House press secretary Jen Psaki are high-stakes settings where the speaker holds significant power, controlling the narrative.
\end{itemize}

\subsection{Analysis}
Informed by the taxonomy of interruption types from prior work and research highlighting the need to tailor interruption handling to the intention (type) of the interruption (detailed in Section \ref{sec:related-work-human-human-interaction}), we coded the videos for 1) intention of the interrupter---either cooperative (\ie agreement, assistance, or clarification) or disruptive---and 2) the strategy individuals use to manage interruptions---yield or hold the floor. Our goal was to identify interaction patterns that reveal how humans handle different types of interruptions. Interaction patterns were created based on the sequence of these two states. One coder transcribed and coded the interruptions, while a secondary coder independently coded $10\%$ to assess inter-coder reliability (Cohen's Kappa $=1$). 



\begin{figure}
\centering
\includegraphics[width=\columnwidth]{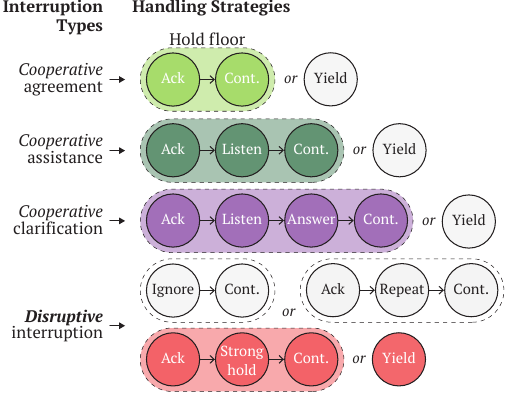}
\caption{Interaction patterns of interruption handling in human conversations. The highlighted strategies were implemented in our system (Fig. \ref{fig:interruption-handling}). \sally{``Ack'' denotes acknowledge and ``Cont.'' denotes continue.}} 
\label{fig:interaction-patterns}
\end{figure}

\subsection{Interaction Patterns}
We identified the following interaction patterns (Fig. \ref{fig:interaction-patterns}):

\subsubsection{Cooperative: agreement $\rightarrow$ hold floor} When speakers encountered cooperative agreement, they typically acknowledged the interruption and continued speaking ($n_D=26$, $n_T=48$, $n_P=2$). A deviation from this pattern was noted in only one case, where the speaker yielded the floor to the interrupter upon agreement during a discussion ($n_D=1$). 
\subsubsection{Cooperative: assistance $\rightarrow$ hold floor} When the interrupter offered assistance, the speaker typically acknowledged it, briefly paused for the interrupter to finish (usually a few words), then resumed speaking ($n_D=7$, $n_T=10$). A deviation occurred a few times, where the speaker yielded the floor to the interrupter ($n_D=4$, $n_T=2$). 
\subsubsection{Cooperative: clarification $\rightarrow$ hold floor or yield} When the interrupter sought clarification, the speaker typically acknowledged the question, paused for the interrupter to finish, answered, and then resumed their original thought ($n_D=5$, $n_T=8$, $n_P=1$). However, in contentious discussions, the speaker sometimes yielded the floor ($n_D=3$, $n_P=1$), showing that clarifying questions can be used to assert alternative points and take control of the floor from the speaker. 
\subsubsection{Disruptive$\rightarrow$ hold floor or yield} 
Disruptive interruptions were the most common (130 of 246) and showed different patterns across tasks. Speakers either held the floor ($n_D=19$, $n_T=37$, $n_P=7$) or yielded it ($n_D=41$, $n_T=17$, $n_P=7$). In debates, speakers often yielded due to the polarized nature of the discussions, while in less polarized settings (talk shows) they typically held the floor, where disruptive interruptions can be more acceptable.

We identified three strategies speakers used to hold the floor: 1) ignore the interruption and continue speaking, 2) acknowledge the interruption while implicitly signaling intent to retain floor by repeating a few words spoken at the time of interruption and continue speaking, and 3) explicitly state intent to continue (\eg \textit{``Let me finish my thought''}) and then continue speaking. The first two strategies were more common for mild interruptions later in the speaker’s turn, after part of their message had been conveyed. The third strategy was used by the speaker to defend their turn from more aggressive disruptive interruptions (\eg disruptive interruption early in their turn or repeated interruption attempts). In multi-party discussions, other participants sometimes intervened to protect the speaker's turn during such aggressive interruptions. Speaker acknowledgments to interruptions were mostly non-verbal (\ie nods and mutual gaze). Verbal acknowledgements (\eg ``sure'' or ``yeah'') occurred in only 13 cases. Similar to prior work \cite{andrist2014conversational}, speakers averted gaze to hold floor, and held prolonged mutual gaze until interrupter took over to yield floor. 

We used these interaction patterns to guide the design of our interruption handling system for conversational robots.


\begin{figure*}
\centering
\includegraphics[width=0.89\textwidth]{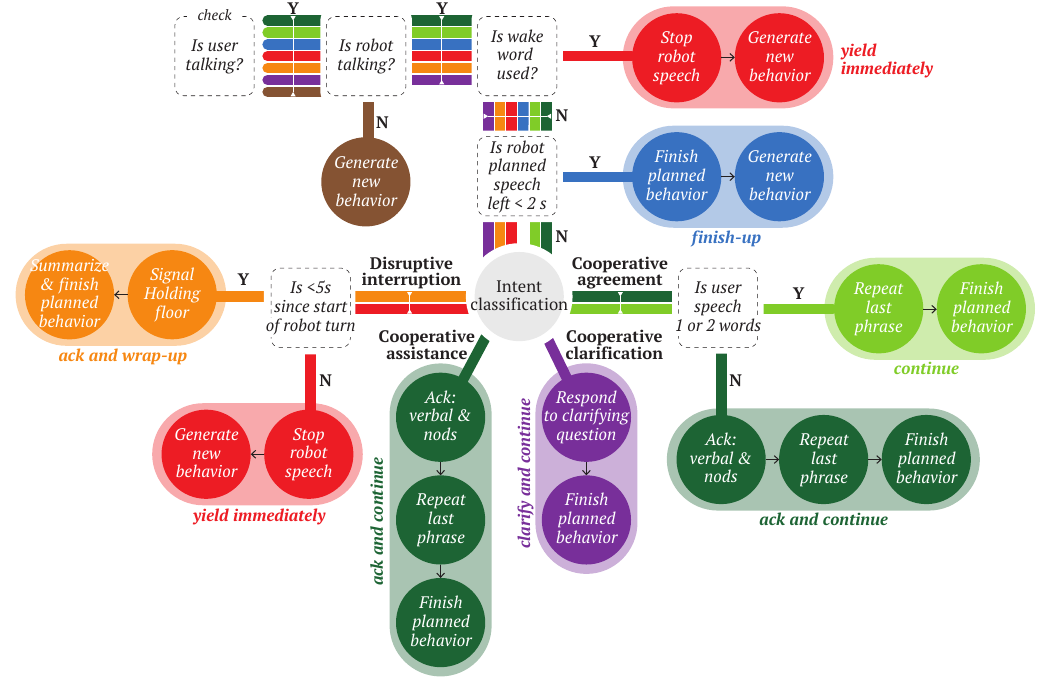}
\caption{\sally{Metro-map-inspired diagram of the interruption handling system, illustrating how user speech flows through the interruption detection, intent classification, and interruption handling modules. It demonstrates how the system selects the handling strategy based on
the predicted user intention given overlapping speech between
the user and the robot. We use ``Line $<$color$>$'' to refer to
different interruption handling paths in the figure. ``Ack'' denotes acknowledge and ``Cont.'' denotes continue.}}
\label{fig:interruption-handling}
\end{figure*}

\section{Interruption Handling System for Conversational Robots}
\label{sec:system}
Our interruption handling system has three main modules: interruption detection, intent classification, and interruption handling. Fig. \ref{fig:interruption-handling} illustrates how user input flows through it. We provide an example implementation of our system integrated into an LLM-powered social robot built on Platform for Situation Intelligence ($\backslash$psi)\footnote{\raggedright \label{footnote1} Supplemental materials (contains additional implementation details): \href{https://intuitivecomputing.github.io/publications/2025-rss-cao-supp.pdf}{https://intuitivecomputing.github.io/publications/2025-rss-cao-supp.pdf}. \\ \phantom{1--} Code: \href{https://github.com/intuitivecomputing/interruption-handling-system}{https://github.com/intuitivecomputing/interruption-handling-system}.}. 

\subsection{User-Initiated Interruptions Detection}
We detect interruptions by monitoring for simultaneous speech. As overlapping speech towards the end of turn is common during speaker floor exchanges and is not considered an interruption \cite{hilton2018does}. So, if at the time of the overlapping speech, the \textbf{robot has less than two seconds of planned speech left}, the system simply ignores it and ``\textbf{finish-up}'' (Line \circlewithoutnumber{customBlue}{4}{1}). Verbal backchannels are also not considered interruptions; however, distinguishing between verbal backchannel and short interruption (\eg ``No'') requires identifying the intent of the interrupter, detailed in the next section. 

Additionally, as users often prefer more control and predictability in human-agent interactions than human-human interactions \cite{clark2019makes}, we added support for explicit wakeword-triggered interruptions. When wakeword (\ie the robot's name) or ``stop'' is heard as overlapping speech, the system always \textbf{yield immediately} (Line \circlewithoutnumber{customRed}{4}{1}). 

\subsection{Interrupter Intention Classification}
In human communication, the speaker's choice of interruption handling strategy depends on the intention of the interrupter. So, we prompt-engineered a large language model (GPT-4o-mini) to classify the intention of the interrupter into cooperative agreement (includes backchannels), cooperative assistance, cooperative clarification, and disruptive interruption given the conversational history and the amount of time elapsed in the turn since the robot started talking. 

Both backchannels and cooperative agreements are used to support the speaker by expressing attention, understanding, and agreement. In this work, we differentiate them by overlap length: short overlaps (one or two words) indicate backchannels, while longer overlaps are cooperative agreements.

\subsection{Interruption Handling}
Following the interaction patterns identified in Section \ref{sec:design}, the system adopts different interruption handling strategies based on the predicted intention of the interrupter: 

\begin{itemize}

\item \textbf{Cooperative Agreement:} If the cooperative agreement is only one or two words, it is likely a verbal backchannel, which the system disregards it and continues with the remaining planned content from the last punctuation mark (\textbf{continue}, Line \circlewithoutnumber{customLightGreen}{4}{1}). If the cooperative agreement contains more than two words, the system acknowledges verbally (with  ``ya'', ``yes'',  ``uhhum'', or ``sure''), nods, and then continues with the remaining planned content (\textbf{ack and continue}, Line \circlewithoutnumber{customDarkGreen}{4}{1}). 

\item \textbf{Cooperative Assistance:} The system acknowledges verbally (with ``yeah'', ``yes'', or ``thanks''), nods, and continues with the remaining planned content from the last punctuation mark. (\textbf{ack and continue}, Line \circlewithoutnumber{customDarkGreen}{4}{1}).

\item \textbf{Cooperative Clarification:} Cooperative clarifications are handled by an LLM prompted to address the clarification requested and then continue with the remaining previously planned content (\textbf{clarify and continue}, Line \circlewithoutnumber{customPurple}{4}{1}).

\item \textbf{Disruptive Interruption:} Disruptive interruptions occurring within 5 seconds of the robot starting to speak are considered aggressive disruptive interruptions by our system. In such cases, an LLM is prompted to generate behavior for the robot to express its intent to maintain the turn (\eg ``Let me finish my thought'' in Fig. \ref{fig:teaser}), provide a summary of the remaining content, and then yield (\textbf{ack and wrap-up}, Line \circlewithoutnumber{customOrange}{4}{1}). For milder disruptive interruptions, the system yields the floor immediately and a new robot response is generated based on the content of the interruption (\textbf{yield immediately}, Line \circlewithoutnumber{customRed}{4}{1}). 
\end{itemize}

\label{sec:methods}
\section{Evaluation Study: Methodology}
We conducted a user study to validate our system by integrating the interruption handling system into an LLM-powered social robot. We implemented basic social robot interaction behavior by hand-crafting a bank of facial expressions (happy, satisfied, excited, interested, surprised, and thinking) and head positions (left gaze, right gaze, look at screen, left nod, right nod, thinking), idle behaviors, gaze aversion, and leveraged an LLM (GPT-4o-2024-05-13) to generate contextualized robot speech and select fitting facial expressions, head movements, and task actions (additional details provided in supplemental materials\footnoteref{footnote1} Section II). 



\subsection{Inducing User-Initiated Interruptions}
\label{sec:inducing-user-initiated-interruptions}
To encourage user-initiated interruptions, we made the following design choices in our implementation: 

\begin{enumerate}
    \item \textit{{Task}}: We contextualized our system in a \textbf{timed decision-making} and a \textbf{contentious discussion} task, anticipating that time pressure and controversy can encourage interruptions. In the decision-making task, participants had five minutes to select seven out of fifteen items to help them survive in a desert survival simulation. The robot was instructed to persuade, rather than simply agree, with users. A countdown timer was shown to add time pressure. In the discussion task, participants prepared for a 2-minute presentation on whether the federal government should abolish capital punishment. The robot remained neutral until the participant clearly stated their position, at which point it adopted an opposing stance to encourage consideration of alternative perspective.

    \item \textit{Robot persona}: 
    We designed the robot to have a feminine name (``Luna"), a feminine voice (Google en-US-Standard-H), gaze aversion while it is talking, and generally positive facial expressions (see supplemental materials\footnoteref{footnote1} Fig. 1). Prior research highlights that gender significantly affects interruption patterns; women are more likely to be interrupted than men, particularly when they avoid direct eye contact with their conversation partners \cite{kennedy1983interruptions}. Additionally, women tend to smile more during conversations, which has been linked to a higher likelihood of being interrupted \cite{kennedy1983interruptions}.


    \item \textit{Pre-programmed events}: 
    \label{sec:pre-programmed-events}
    We designed two pre-programmed events\footnotemark[1] per task where the robot holds the floor for roughly one minute to provoke user-initiated interruptions. In event one, triggered at the start of the discussion, the robot introduces itself and provides disclaimers. In event two, triggered 3.5 minutes into the discussion, the robot tells an unrelated fun story or facts.
\end{enumerate}


\subsection{Study Procedure}
\sally{We set up two cameras to record the participants’
interactions with the robot from both the front-view and back-view during the study (see Fig. \ref{fig:-example-conversation-diagram}).} After providing consent, participants completed a demographic questionnaire and watched a 1-minute sample interaction video. They then engaged in a practice task, designed to familiarize them with the robot, answering four factual questions about space exploration. Following the practice task, participants completed two tasks (in random order) and completed a post-task questionnaire after each task. The study concluded with a semi-structured interview aimed to understand their overall experience working with the robot and to collect feedback on robot interruption handling.

\subsection{Participants}
We recruited 21 participants (11 female, 10 male), aged 18 to 30 (M=$22.38$, SD=$3.58$), through convenience sampling from the local community via electronic newsletters and mailing lists. Most participants had limited experience using speech-based AI technology (M=$2.81$, SD=$1.25$, Min=$1$, Max=$5$, 5-point Likert scale with 1 being no experience and 5 being extensive experience) and reported that they have only used it for simple commands, \ie setting alarms and checking the weather. The study took roughly 45 minutes, and participants were compensated at $\$15.00$ per hour. The study was approved by our institutional review board (IRB).

\subsection{Metrics}
We constructed two subjective scales (see supplemental materials\footnoteref{footnote1} Section III for details) from the post-task questionnaire to explore how interruption handling might affect the user's:
\begin{itemize}
    \item \textit{Perceived inclusion in discussion} (rating scale 1--5): Three-item scale (Cronbach's $\alpha = .78$) measuring how heard, respected, and valued users felt during the discussion.
    \item \textit{Overall satisfaction with discussion} (rating scale 1--5): Nine-item scale (Cronbach's $\alpha = .77$) assessing user's enjoyment, learning, contentment with outcomes of the discussion, and willingness to engage in future discussions.
\end{itemize}

\subsection{Coding}
One coder transcribed all interactions and extracted conversation snippets with overlapping speech. For ground truth and fine-grained understanding of interruptions, two coders independently analyzed 100\% of the interruptions, classifying them by type (disruptive, cooperative agreement, assistance, or clarification) and by attributes such as statement (opinion or non-opinion), question (opinion or factual), or verbal back-channel, and assessed whether each interruption was successfully addressed, meaning that the robot effectively responded to the interrupter without causing a conversation breakdown. Cohen's kappa was $0.92$, and coders resolved conflicts through discussion.

\begin{figure*}[h]
\centering
\includegraphics[width=\textwidth]{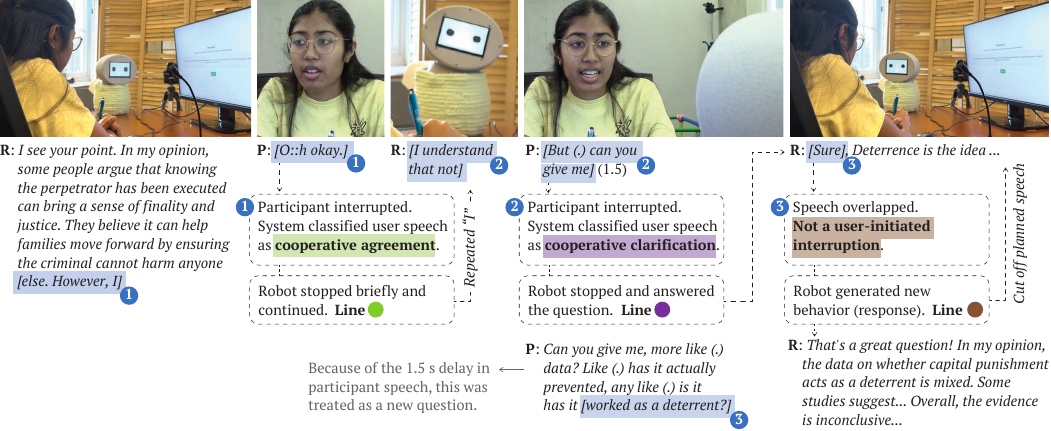}
 \caption{Example conversation where the system handles various types of compounded interruptions. Overlap \circlewithnumber{customBlue}{3}{1} is an example of a robot-initiated interruption. We used the Jeffersonian transcription system \cite{park2022benefits}. \sally{``R'' denotes robot and ``P'' denotes participant.}}

\label{fig:-example-conversation-diagram}
\end{figure*}

\label{sec:results}
\section{Evaluation Study: Findings}
We identified $121$ user-initiated interruptions from $206.59$ minutes of interaction data (\eg Fig. \ref{fig:-example-conversation-diagram})
involving $525$ user query-robot response pairs, or turns (\dm: $312$ turns, \disc: $213$ turns). The timed decision-making task ($n=86$) 
had more than twice as many user-initiated interruptions as the contentious discussion task ($n=35$).


\begin{table}[h]
\centering
\caption{Types of interruptions observed and examples.}
\small
\resizebox{\columnwidth}{!}{%
\begin{tabular}{p{1.45cm}|p{1.1cm}|M|p{0.2cm}}
    \textbf{Speech act} & \textbf{Attribute} & \normalfont \normalsize \textbf{Example} & \textbf{n}  \\
    \Xhline{3\arrayrulewidth}
    \multicolumn{4}{l}{\textbf{Cooperative agreement} ($n=18$) } \\
    \hline
    \multicolumn{2}{l|}{Verbal back-channel} & ``Yeah'', ``Okay'' & 13 \\
    \cline{1-4}
    \multirow{2}{*}{Statement} & Opinion & ``That's a good idea'' & 2 \\
    \cline{2-4}
                               & Non-opinion &  ``Alright, I'm taking your suggestions'' & 3 \\
    \Xhline{3\arrayrulewidth}
    \multicolumn{4}{l}{\textbf{Cooperative assistance} ($n=2$) }\\
    \hline
    Statement & Opinion & ``Another thing that I was  thinking was jack knife.''& 2 \\
    \Xhline{3\arrayrulewidth}
    \multicolumn{4}{l}{\textbf{Cooperative clarification ($n=9$)} }  \\
    \hline
    \multirow{2}{*}{Question} & Opinion & ``Uh, do we need the raincoat and the parachute?'' & 5 \\
    \cline{2-4}
                               & Factual &  ``What percent?'' & 4 \\  
    \Xhline{3\arrayrulewidth}
    \multicolumn{4}{l}{\textbf{Disruptive interruption ($n=92$)} }  \\
    \hline
    \multirow{2}{*}{Question} & Opinion & ``Luna, what do you think  about adding the pistol to the list?''& 42 \\
    \cline{2-4}
                               & Factual &  ``How many states have  capital punishment?'' & 16 \\  
    \cline{1-4}
    \multirow{2}{*}{Statement} & Opinion & ``I do not think that the  cosmetic mirror is necessary.  I think we should change it into 45 caliber pistol'' & 7 \\
    \cline{2-4}
                               & Non-opinion &  ``Uh, Luna we we don't  have time'' & 27 \\  
    \Xhline{3\arrayrulewidth}
\end{tabular}
}
\label{tab:interruption-classification}
\end{table}

\subsection{The majority of interruption attempts were disruptive}
Based on the interruption intent coded by the research team, the most common type of interruption was disruptive ($n=92$, $76.03\%$), followed by cooperative agreement ($n=18$), cooperative clarification ($n=9$), and rarely, cooperative assistance ($n=2$). Participants attempted disruptive interruptions in various ways, including asking opinion ($n=42$ out of $92$, $45.65\%$) and factual ($n=16$) questions or issuing opinion ($n=7$) and non-opinion ($n=27$) statements. See Table \ref{tab:interruption-classification} for more details and example interruptions. 

\subsection{$88.78\%$ of interruption intent was correctly classified}

Five interruptions triggered pre-programmed events (see description in Sec. \ref{sec:pre-programmed-events}), bypassing the interruption handling system. During a 35-second time frame in one study, there was an abnormally large delay in the speech-to-text API, likely caused by a network error, causing the system to be unable to detect user speech. Hence, our system did not handle the five interruptions (all disruptive) during this time. We exclude these ten cases from the rest of our analysis. 

Amongst the interruptions handled by the system ($n=111$), participants used wakeword(s) in $47$ interruptions (\dm: $n=36$, \disc: $n=11$). Our system is designed to immediately yield when a wakeword is ``heard'' during overlapping speech ($n=43$). It treats all interruption attempts containing wakeword(s) as strong disruptive interruptions without passing the intent classifier module (SR2, Line \circlewithoutnumber{customRed}{4}{1} in Fig. \ref{fig:interruption-handling}). Due to speech recognition errors, wakewords were not ``heard'' in $4$ out of $47$ cases. Additionally, interruptions made when less than two seconds of robot-planned speech remained ($n=13$) also bypassed the intent classifier module (SR3, Line \circlewithoutnumber{customTeal}{4}{1} in Fig. \ref{fig:interruption-handling}). 
Of the $55$ interruption attempts that passed through the intent classification module\footnote{$111$ total handled interruptions minus $43$ with wakeword(s) detected, and $13$ occurring near the end of robot-planned speech.}, the module correctly classified the intent of $44$ interruptions ($80.00\%$). Overall, including interruptions with wakeword(s) detected ($n=43$), our system incorrectly classified the intent in $11$ out of $98$ ($11.22\%$) interruptions, meaning $88.78\%$ were correctly classified. 

\subsection{$93.69\%$ of the interruptions were successfully handled}
Not all interruption intent classification errors resulted in unsuccessful handling, nor were all correctly classified interruptions handled successfully. Of the $111$ interruptions handled, $107$ cases were the user's first interruption attempt during that turn, and $100$ out of the $107$ ($93.46\%$) successfully handled. Among the seven unsuccessful initial attempts, the robot continued speaking in four cases due to intent classification errors, leading the users to re-attempt. All four ($100.00\%$) re-attempts were successful. Overall, $104$ out of $111$ ($93.69\%$) interruptions were successfully handled.

\footnotetext[3] {Two cooperative agreements, one assistance, one clarification, and nine disruptive interruptions---two used wakewords but were not ``heard'' by system due to speech recognition error
---were handled through finish-up.}
\footnotetext[4]{LLM failed to generate correctly formatted robot behavior, so the robot performed default LLM error handling. See supplemental materials\footnoteref{footnote1} Section II for details.}

\begin{table}[h]
\caption{Types of interruptions and handling strategies (Misclassified interruptions are highlighted).}
\small
\begin{tabular}{p{0.7cm}|p{1.4cm}|p{2.7cm}|p{0.15cm}|p{0.8cm}}
\multicolumn{1}{l|}{
\begin{tabular}[l]{@{}l@{}}\textbf{Interruption}\\ \textbf{type (coded)}\end{tabular}}   
& \begin{tabular}[l]{@{}l@{}}\textbf{Type} \\ \textbf{(by model)}\end{tabular} & \multicolumn{1}{l|}{\begin{tabular}[l]{@{}l@{}}\textbf{Interruption} \\ \textbf{handling}\end{tabular}} & \multicolumn{1}{c|}{\textbf{n}} & \begin{tabular}[l]{@{}l@{}}\textbf{$\%$} \\ \textbf{Success}\end{tabular}     
\\ 
\Xhline{3\arrayrulewidth}
\begin{tabular}[c]{@{}l@{}}\textless 2s of plan-\\ ned speech \end{tabular} & N/A & Finish-up\footnotemark[3] & $13$ & $100\%$\\ 
\Xhline{3\arrayrulewidth} & & Continue & $12$ & $91.67\%$\\
& \multirow{-2}{*}{Agreement} & Ack and continue & $2$ & $100\%$\\
\multirow{-3}{*}{\begin{tabular}[c]{@{}l@{}}\textbf{Agreement} 
\end{tabular}}    
& \cellcolor[HTML]{DEDCDC}Disruptive                                                                                    & \cellcolor[HTML]{DEDCDC}Yield immediately & \cellcolor[HTML]{DEDCDC}$2$ & \cellcolor[HTML]{DEDCDC}$100\%$\\ 
\Xhline{3\arrayrulewidth}
\begin{tabular}[c]{@{}l@{}}
\textbf{Assistance}\end{tabular} & Assistance  & Ack and continue & $1$ & $0\%$\\ 
\Xhline{3\arrayrulewidth} & Clarification & Clarify and continue & $7$ & $100\%$\\
\multirow{-2}{*}{\begin{tabular}[c]{@{}l@{}}\textbf{Clarification} 
\end{tabular}} & \cellcolor[HTML]{DEDCDC}Disruptive                                                                                   & \cellcolor[HTML]{DEDCDC}Yield immediately & \cellcolor[HTML]{DEDCDC}$1$ & \cellcolor[HTML]{DEDCDC}$0\%$ \\ 
\Xhline{3\arrayrulewidth} & & Yield immediately & $17$ & $100\%$ \\
& \multirow{-2}{*}{Disruptive} & Ack and wrap-up & $5$ & $100\%$ \\
& \cellcolor[HTML]{DEDCDC} Agreement& \cellcolor[HTML]{DEDCDC}Ack and continue & \cellcolor[HTML]{DEDCDC}1 & \cellcolor[HTML]{DEDCDC}$0\%$ \\
& \cellcolor[HTML]{DEDCDC} Assistance& \cellcolor[HTML]{DEDCDC}Ack and continue & \cellcolor[HTML]{DEDCDC}1 & \cellcolor[HTML]{DEDCDC}$0\%$ \\
& \cellcolor[HTML]{DEDCDC} Clarification & \cellcolor[HTML]{DEDCDC}Clarify and continue & \cellcolor[HTML]{DEDCDC}3 & \cellcolor[HTML]{DEDCDC}$100\%$ \\
\multirow{-6}{*}{\begin{tabular}[c]{@{}l@{}}\textbf{Disruptive} \\ \\ wakeword \\ not used 
\end{tabular}}   & 
\cellcolor[HTML]{DEDCDC} 
Clarification & \cellcolor[HTML]{DEDCDC}LLM error handling\footnotemark[4]      & \cellcolor[HTML]{DEDCDC}1 & \cellcolor[HTML]{DEDCDC}$100\%$ \\ 
\hline & N/A  & Yield immediately  & $43$ & $100\%$ \\
\multirow{-2}{*}{\begin{tabular}[c]{@{}l@{}} wakeword \\ used\end{tabular}}  & \cellcolor[HTML]{DEDCDC} Agreement & \cellcolor[HTML]{DEDCDC}Continue & \cellcolor[HTML]{DEDCDC}2 & \cellcolor[HTML]{DEDCDC}$0\%$ \\    
\Xhline{3\arrayrulewidth}
\end{tabular}
\label{tab:interruption-handling}
\end{table}   

\subsection{Speech recognition errors caused the majority of interruption handling failures}
To better understand the cause of unsuccessfully handled interruptions ($n=7$), we analyze their potential links to intent classification errors ($n=11$), as shown in Table \ref{tab:interruption-handling}.

\begin{itemize} 
    \item \textit{Misclassified disruptive interruption as cooperative agreement ($n=3$).} 
    The robot continued speaking due to misclassification; two of which contained a wakeword but were still misclassified due to speech recognition error. 

    \item \textit{Misclassified disruptive interruption as cooperative assistance ($n=1$).} 
    The participant attempted to interrupt with \textit{``The first aid kit''} but abandoned the interruption mid-sentence, which the system interpreted as cooperative assistance. Five seconds later, the participant reattempted a disruptive interruption, and the robot yielded immediately. 
    
    \item \textit{Misclassified disruptive interruption as cooperative clarification ($n=4$).} 
    Despite the miss-classification, these interruptions were considered successfully handled, as the robot answered the user's questions. 

    \item \textit{Misclassified cooperative agreement as disruptive ($n=2$).}
    While the robot yielded to cooperative agreements (\eg \textit{``Oh, go ahead and keep going''}), these cases did not manifest as unsuccessful interruption handling.

    \item \textit{Misclassified cooperative clarification as disruptive ($n=1$).} 
    A clarifying question was misclassified as disruptive and inadequately answered due to speech recognition error. 
\end{itemize}

Although correctly classified, two interruptions were unsuccessfully handled due to errors in determining where the robot left off at the time of interruption ($n=2$). For example, 
\begin{quote}
\fontfamily{lmr}\selectfont\small\textcolor[gray]{0.2}{}
\textbf{robot:} Great! The flashlight is now on our list. 

\phantom{robot:} Next, I suggest [a] \makebox[3.7cm][r]{[overlapping speech]}

\textbf{\phantom{r}user:} \phantom{Next, I suggest} [okay]

\textbf{robot:} It can be used as a shelter and for signaling.

\textbf{\phantom{r}user:} what was it?

\textbf{robot:} I suggested a parachute. It can serve \ldots
\end{quote}
Since text-to-speech services do not provide word-by-word timestamps, the system estimates what the robot has said based on the average speaking rate and the amount of time elapsed. To prevent missing context, we also designed the robot to repeat its speech up to the last punctuation mark. 
Despite this, misalignment still occurs; however, users were able to recover by asking a follow-up question about the missing information in both cases.


\subsection{Unsuccessful interruption handling led to decreased perceived inclusion, and lower discussion satisfaction.}
While unsuccessful interruption handling was relatively rare, we conducted exploratory analyses to see how the interruption handling error might affect users' perception of the discussion. Pearson correlation coefficient was used to examine the linear relationship between the number of unsuccessfully handled interruptions, user's perceived inclusion in the discussion, and user's satisfaction with the discussion. 

We found a negative correlation between the number of unsuccessfully handled interruptions and the user's perceived inclusion in the discussion, $\rho(42)=-.43, p=.005$. We also found a negative correlation between the number of unsuccessfully handled interruptions and the user's satisfaction with the discussion, $\rho(42)=-.35, p=.021$. 

Nonetheless, based on the post-study interviews, participants generally perceived their experience with the robot positively ($n=15/21$, describing it as ``good'', ``nice'', or ``helpful''). Among those who recalled initiating interruptions ($n=18$), 14 found it ``easy'' to interrupt the robot or were generally satisfied with how it handled interruptions, while four found it ``hard''---one because the system failed to detect five interruptions due to network error, and three because they wanted the robot to always yield immediately.

\label{sec:discussion}
\section{Discussion}

In this work, inspired by interaction patterns observed in human conversations, we designed and implemented an interruption handling system for conversational robots. \sally{To the best of our knowledge, no existing robotic system has
integrated intention classification into its interruption handling
framework. Most existing conversational robots either ignore
interruptions or always yield to any overlapping speech. Systems that ignore interruptions are prone to conversational
breakdowns as they disregard the user’s intention to shift the
conversation. Alternatively, systems that always yield break
the interaction flow when they yield to non-disruptive overlapping speech (\eg cooperative agreement, cooperative assistance, self-talk, and backchannels), which can result in user
frustration over time. The distribution of types of interruptions
depends on the task context (\ie a higher number of disruptive
interruptions in our survival task, likely due to the time
constraint). In our study, a system that ignores interruptions
would fail to handle cooperative clarification (n = 8) and
disruptive interruptions (n = 73), leading it to successfully
handle only 27.03\% of the interruptions. A system that always
yields would fail to handle cooperative agreement (n = 16)
and cooperative assistance (n = 1), causing it to successfully
handle 84.68\% of the interruptions. In comparison, our proposed intention-based system successfully handled 93.69\% of
the interruptions. This shows the benefit of intention-based
interruption handling.} Next, we discuss design implications based on insights from our exploration to help guide future designs of interruption handling behavior in conversational robots.

\subsection{Robot's Role, Task Context, and Interruption Handling}
While floor holding is a common strategy for handling interruptions in human conversations, robots holding the floor---particularly during disruptive interruptions (\eg Fig. \ref{fig:teaser})---were not perceived favorably by all participants. In fact, one participant explained that \quotes{``I felt like I am trying to solve the issue, and I am just using you [the robot] as a tool to help me,''} so their ideal interaction is \quotes{``the moment I speak, it wants to just like immediate listen, and like my speech should take precedence over anything it says''}. 
The participants perceived the robot's role in the task as assistive rather than collaborative, so they expected it to yield immediately at every turn. This shows the importance of aligning the robot's role and task context with its interruption handling behavior. Future research should explore how a robot's designed and perceived role influences users' expectations for how the robot should handle interruptions.

On the other hand, interruption handling design may help reinforce a robot's intended role in a given context. For example, an authoritative robot in a critical task should hold the floor more aggressively to help establish the power dynamics, while a robot designed to be an assistive tool in a causal task should adopt a more flexible and accommodating interruption-handling approach. However, this needs to be approached with caution, as studies showed that people are prone to complying with authoritative robots even if it is wrong \cite{karli2023if}. Future work is needed to define the balance between when a robot should hold the floor in conversations based on its role and the task. 

\subsection{Breaking Habits: From Scripted to Natural HRI}
While some participants preferred the predictability of having a ``designated stop mechanism'' (wakewords), some also felt uncomfortable using the robot's name ( ``Luna'') or ``stop'' to initiate interruptions, as it can be viewed as rude in human-human conversations. However, one participant, who reported to use voice assistants on a daily basis, began nearly every query with ``Luna''. When they initiated interruptions, they consistently said ``Luna'' and waited for the robot to stop speaking before proceeding with their query. While their conversation with the robot was fast-paced, it was very rigid. This was not a unique case in our study. The habits users have developed when interacting with commercial voice assistants constrain their interactions with systems capable of more natural, conversational exchanges---such as ours that can handle interruptions more fluidly. It may also limit their perception of the robot's role to that of an assistive query-answering tool. Future research should design more natural yet predictable interactions and explore the lasting influence of current technology use on how users engage with emerging technologies that offer more advanced conversational capabilities.


\subsection{Limitations}
\sally{We designed our interruption handling framework based on observations from human-human conversation data from YouTube. However, the sample size was small and limited to a few specific settings. Additionally, our HRI study focused solely on robot as cognitive aid in timed decision-making and contentious discussion scenarios. Future work is needed to explore interruption handling in physical tasks, as well as how interruption handling can be personalized and tailored based on user behavior (\eg repeated interruptions), user preferences (\eg their perceived role of the robot), and task context (\eg time constraint) for designing more comprehensive interruption management for HRI. Furthermore, our study used convenient sampling and did not design interruption handling to support special population. Future work should explore how interruption handling may be designed to improve interactions for special populations (\eg older adults with neurodegenerative
disease). Finally,} our evaluation was limited to one-time interactions in the lab setting. Future work should explore longer-term, multi-session interactions in the wild. 

\sally{Our system also had a few limitations. As shown in our study, speech recognition errors were the primary cause of failure for the intention classification module. While the LLM absorbed some speech recognition errors (some errors did not affect the LLM’s ability to understand context), our current system cannot detect discrepancies between the transcribed speech and actual user speech. Future work could investigate using multimodal inputs with multimodal LLMs as an additional information source.} Additionally, our system does not account for non-verbal interruptions. We observed seven instances where the users' body language (\eg opening their mouths) indicated an intention to interrupt the robot without making any sounds. Future work should explore how non-verbal interruption can be leveraged in interruption handling and how non-verbal interruptions should be handled by conversational robots. In addition, our system focuses on interruption handling in dyadic conversations; future work should explore how robots should handle interruptions in multi-party and team situations. 


\section*{Acknowledgments}
\sally{This work was supported by the National Science Foundation award \#2141335.}

\sally{\textbf{AI Statement}. This paper has been proofread by a language model. The authors verified that the resulting content accurately reflects their original intent.}

\sally{\textbf{CRediT author statement}. Shiye Cao: Conceptualization, Methodology, Software, Validation, Formal analysis, Investigation, Data curation, Writing, Visualization, Project Administration. Jiwon Moon: Conceptualization, Methodology, Software, Validation, Formal analysis, Investigation, Data curation, Writing, Visualization. Amama Mahmood: Conceptualization, Methodology, Formal analysis, Investigation, Data curation, Writing, Visualization. Victor Nikhil Antony: Methodology, Software, Writing. Ziang Xiao: Conceptualization, Methodology, Writing-Review and editing. Anqi Liu: Conceptualization, Methodology, Writing-Review and editing. Chien-Ming Huang: Conceptualization, Methodology, Writing-Review and editing, Visualization, Supervision, Funding acquisition.}

\balance

\newpage

\bibliographystyle{plainnat}
\bibliography{references}

\end{document}